\documentclass[conference]{IEEEtran}
\usepackage{cite}

 \usepackage[pdftex]{graphicx}
\ifCLASSINFOpdf
   \usepackage[pdftex]{graphicx}
\else
\fi
\usepackage[cmex10]{amsmath}
\usepackage{stfloats}

\usepackage{mdwmath}
\usepackage{mdwtab}

\usepackage[tight,footnotesize]{subfigure}

\usepackage[font=footnotesize]{subfig}
\usepackage{stfloats}
\usepackage{amssymb}
\begin{document}
%
\title{The role of magnetic fields \\
in star formation\\
}

\author{\IEEEauthorblockN{Raquel Salmeron}
\IEEEauthorblockA{Research School of Astronomy \& Astrophysics \\ Research School of Earth Sciences \\
Planetary Science Institute\\
The Australian National University\\
Cotter Road
Weston Creek, ACT 2611
AUSTRALIA\\
Email: raquel@mso.anu.edu.au}
}


%


\maketitle

\begin{abstract}
Star formation is thought to be triggered by the gravitational collapse of the dense cores of molecular clouds. Angular momentum conservation during the collapse results in the progressive increase of the centrifugal force, which eventually halts the inflow of material and leads to the development of a central mass surrounded by a disc. In the presence of an angular momentum transport mechanism, mass accretion onto the central object proceeds through this disc, and it is believed that this is how stars typically gain most of their mass. However, the mechanisms responsible for this transport of angular momentum are not well understood. The most promising are \emph{turbulence viscosity} driven by the magnetorotational instability (MRI), and \emph{outflows} accelerated centrifugally from the surfaces of the disc. Both processes are powered by the action of magnetic fields and are, in turn, likely to strongly affect the structure, dynamics, evolutionary path and planet-forming capabilities of their host discs. The weak ionization of protostellar discs, however, may prevent the magnetic field from effectively coupling to the gas and drive these processes. Here I examine the viability and properties of these magnetically driven processes in protostellar discs. The results indicate that, despite the weak ionization, the field is able to couple to the gas and shear for fluid conditions thought to be satisfied over a wide range of radii in these discs.  
\end{abstract}


%
\IEEEpeerreviewmaketitle

\section{Introduction}
The current paradigm for star formation, framed originally by Kant and 
Laplace in the $18^{\rm th}$ century, suggests that stars are born via gravitational collapse of the dense cores of molecular clouds, in turn condensed out of the diffuse 
interstellar medium.  
During this phase of the star formation process, conservation of angular momentum 
results in the progressive increase of the centrifugal force, which eventually halts the 
infalling gas and leads to the development of a central mass (i.e.~a `protostar') surrounded by a flattened disc of material (an `accretion disc').  The difficulty in progressing past this stage is clear when we recall that this disc is dynamically stable.   We are forced then to address the following question: How does nature overcome this orbital stability, if stars are to form?

A number mechanisms have been invoked over the years to explain this process (e.g.~see the review by Stone et al.~2000 \cite{2000prpl.conf..589S}), but have all proved to be either extremely inefficient (i.e.~molecular viscosity), not general enough (i.e.~the presence of a companion star) or not to work at all (i.e.~convection). In recent times it has been realised that the 
complex interaction of gas and magnetic fields present in the disc may be the missing piece in this puzzle (e.g.~Shakura \& Sunyaev 1973 \cite{1973A&A....24..337S}, Balbus \& Hawley 1998 \cite{1998RvMP...70....1B}). 
Most analytical and numerical studies conducted so far have adopted a number of simplifications to treat the fluid, but they are poor approximations to the gas in the cold, dense `protostellar discs' that surround the forming stars. Angular momentum lies at the core of disc dynamics and to understand how it is transported it is crucial to treat adequately the slippage between the magnetic field and the gas (e.g.~the \emph{magnetic diffusivity}) in these environments. 

\section{Magnetic activity of protostellar discs}
\label{sec:B}

Magnetic fields are thought to play key roles in the dynamics and evolution of protostellar discs. They redistribute angular momentum via  
magnetohydrodynamic (MHD) turbulence (Stone et al.~2000 \cite{2000prpl.conf..589S}) and by magnetocentrifugal winds accelerated from the disc surfaces (and also possibly from the star/disc magnetosphere; e.g.~see the reviews by K\"onigl and Pudritz \cite{2000prpl.conf..759K} and Pudritz et al. \cite{2007prpl.conf..277P}). Additionally, magnetically driven mixing strongly influences the chemistry of the disc (e.g. Semenov et al. 2006 \cite{2006ApJ...647L..57S}, Ilgner \& Nelson 2006 
\cite{2006AA...445..223I}) and critically affects the dynamics, aggregation and overall evolution of dust particles (Turner et al.~2006 \cite{2006ApJ...639.1218T}, Ciesla 2007 \cite{2007ApJ...654L.159C}), the assembly blocks of planetesimals according to the `core accretion' model of planet formation (Pollack et al. 1996 \cite{1996Icar..124...62P}). Furthermore, magnetic fields can also modify the response of the disc to the gravitational perturbations introduced by forming planets and therefore, their rate and direction of migration through the disc (Terquem 2003 \cite{2003MNRAS.341.1157T}, Fromang et al.~2005 \cite{2005MNRAS.363..943F}, Muto et al. 2008 \cite{2008ApJ...679..813M},   Johnson et al.~2006 \cite{2006ApJ...647.1413J}). Jets accelerated via magnetic stresses may be the sites of \emph{chondrule} formation (primitive, thermally processed pieces of rock found in meteorite samples; see e.g.~\cite{2005ASPC341..215C}) and magnetically-driven activity near the disc surface can produce a hot, tenuous corona (e.g. Fleming \& Stone 2003 \cite{2003ApJ...585..908F}) and influence the observational signatures of these objects. In protoplanetary discs, however, the magnetic diffusivity can be severe enough to limit -- or even suppress -- these processes. The particular role magnetic fields are able to play in these environments is, therefore, largely determined by the degree of coupling between the field and the neutral gas. This important topic is discussed next.

\subsection{Magnetic diffusivity}
\label{subsec:Magdiff}

 \begin{figure}[tbp]
\centering
\includegraphics[width=3.0in]{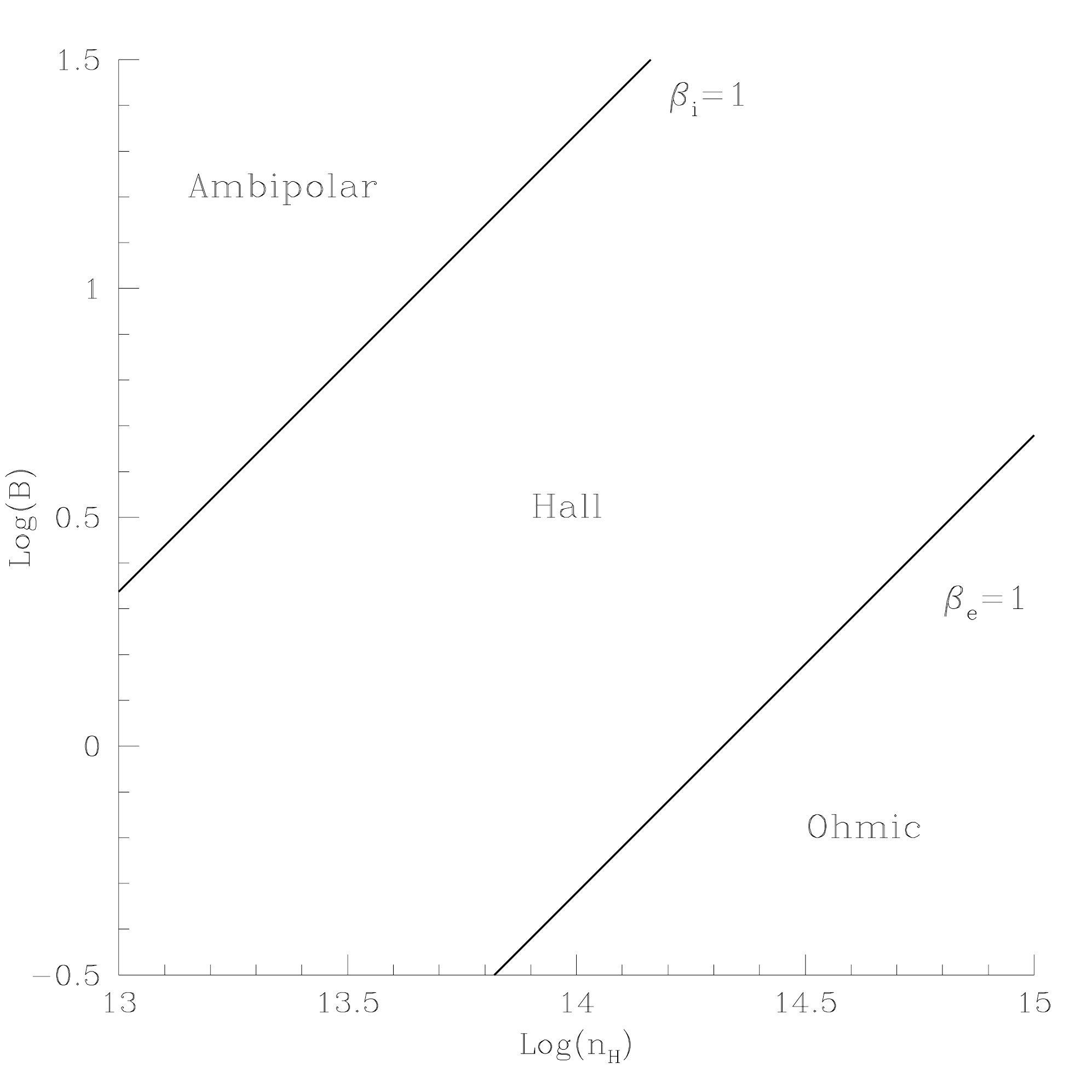}
\caption{Magnetic diffusivity regimes -- ambipolar, Hall and Ohmic --  in a log $n_{\rm H}$ [cm$^{-3}$] -- log $B$ [G] plane for $T = 280$ K. The different regions are delineated by the values of the ion (subscript $i$) and electron (subscript $e$) Hall parameters $\beta_{\rm i}$ and $\beta_{\rm e}$. The $\beta_j$ measure the ratio of the gyrofrequency and the collision frequency of charged species $j$ with the neutrals (in turn a measure of the relative importance of the Lorentz and drag forces on the motion of the charged species). Note that in a minimum--mass solar nebula disc (Weidenschilling 1977 \cite{1977Ap&SS..51..153W}, Hayashi 1981 \cite{1981Hayashi}), the midplane number density at $R = 1$ AU would be $\sim 6 \times 10^{14}$ cm$^{-3}$. This implies that the gas at this location would be in the Ohmic diffusivity regime for field strengths $ \lesssim 2.5$ G and in the Hall limit for stronger fields. }
\label{fig:diff}
\end{figure}

The electrical conductivity of a fluid is determined by the abundance and drifts of the charged species (in general ions, electrons and charged dust grains) embedded in it. The abundance of the charged species (measured by the ionisation fraction) results, in turn, from the balance of ionisation and recombination processes acting in the disc. In protostellar systems, in particular, ionising agents outside the central $0.1$ AU around the protostar are non-thermal (mainly driven by stellar X-ray and UV radiation and interstellar cosmic rays; see e.g.~Hayashi 1981 \cite{1981Hayashi}, Semenov et al.~2004 \cite{2004A&A...417...93S}, Glassgold et al.~2005 \cite{2005ApJ...621..808G}). As a result, in the dense, cold inner regions of discs (perhaps with the exception of the surface layers) the fractional ionisation may not be enough to provide good coupling between the charged and neutral components of the fluid. The degree of coupling between the charged species and the neutral gas is typically measured by the Hall parameter $\beta_j$, the ratio of the gyrofrequency and the collision frequency of species $j$ (taken here to be either ions or electrons and denoted by subscripts $i$ and $e$, respectively), with the neutrals. This parameter measures the relative importance of the  Lorentz and drag forces on the motion of the charged species. In these conditions, the relative drifts of different charged species with respect to the neutrals delineate three different \emph{conductivity regimes}:

\begin{itemize}
\item \emph{Ambipolar diffusion limit}. In this regime, which  is important at relatively low densities, most charged species are primarily tied to the magnetic field through electromagnetic stresses (or $\beta_{\rm e} \gg \beta_{\rm i} \gg 1$).  This implies that the magnetic field is effectively frozen into the ionised component of the fluid and drifts with it through the neutrals. In protostellar disks, this regime is expected to be dominant at radial distances beyond $\sim 10$ AU and near the surface closer in.
\item \emph{Resistive (Ohmic) limit}, in which the ionised species are mainly tied to the neutrals via collisions ($1 \gg \beta_{\rm e} \gg \beta_{\rm i}$). This regime is predominant in high density regions, typically close to the midplane in protostellar discs, where the collision frequency of the charged species with the neutrals is high enough to suppress the drift of the former through the bulk of the gas.  
\item \emph{Hall limit}, characterised by a varying degree of coupling of different charged species with the neutrals. This  regime is important at intermediate densities, in-between those at which the ambipolar diffusion and resistive limits are dominant. In this limit, some charged species (typically electrons) are still tied to the magnetic field whereas more massive particles (such as ions and charged dust grains) have already decoupled to it and are collisionally tied to the neutral gas ($\beta_{\rm e} \gg 1 \gg \beta_{\rm i}$). When this regime dominates, the magnetic response of the disc is no longer invariant under a global reversal of the magnetic field polarity (e.g.~Wardle \& Ng 1999 \cite{1999MNRAS.303..239W}). The Hall regime is expected to dominate over fluid conditions satisfied in large regions in protostellar discs (e.g. Sano \& Stone 2002a,b \cite{2002ApJ...570..314S})
\end{itemize}

In the collisionally-dominated resistive limit, the resulting electrical conductivity is a scalar, the familiar Ohmic resistivity. This is simply because collisions occur in all directions, and the impulse they communicate to the charges is randomly oriented. In the other limits, however, where at least some charged species are well tied to the magnetic field, the conductivity is a tensor (Cowling 1976 \cite{Cowling1976}, Norman \& Heyvaerts 1985 \cite{1985A&A...147..247N}, Nakano \& Umebayashi 1986 \cite{1986MNRAS.218..663N}, Wardle 1999 \cite{1999MNRAS.307..849W})\footnote{Specifically, the conductivity is a tensor whenever the gyrofrequency of some charge carriers is larger than their frequency of collisional exchange of momentum with the neutrals.}. 
This is expected, and reflects the anisotropy that the field is able to impart on the drift of the charges when at least some of them are well coupled to it. 

Previous work on the magnetic activity of discs (e.g. Wardle 1999 \cite{1999MNRAS.307..849W}, Sano \& Stone 2002a,b \cite{2002ApJ...570..314S}, \cite{2002ApJ...577..534S}, Salmeron \& Wardle 2003 \cite{2003MNRAS.345..992S}, 2005 \cite{2005MNRAS.361...45S} and 2008 \cite{2008MNRAS.388.1223S}) have highlighted the importance of incorporating in these studies all the three field-matter diffusion mechanisms described above, as their relative importance is a strong function of location in the disc. This is illustrated in Fig.~\ref{fig:diff}, which shows the regions where the ambipolar, Hall and Ohmic diffusivity terms dominate in a Log $B$ -- Log $n_{\rm H}$ plane. Note that ambipolar diffusion is dominant at low densities and strong fields whereas the opposite is true for the Ohmic regime. The vast, intermediate region in parameter space between these two limits is dominated by the Hall diffusivity.  For example, for $R = 1$ AU, the number density at the disc midplane, according to the minimum--mass solar nebula model\footnote{The surface density of a minimum-mass solar nebula disc is estimated by incorporating sufficient hydrogen and helium to the solid bodies of the solar system to recover solar abundances, and then smoothing the resulting material in a disc-like shape. This model, therefore, expresses the minimum amount of matter required to build the solar system.} (Weidenschilling 1977 \cite{1977Ap&SS..51..153W}, Hayashi 1981 \cite{1981Hayashi}) is $\sim 6 \times 10^{14}$ cm$^{-3}$. This implies that the gas at this location would be in the Ohmic diffusivity regime for field strengths $ \lesssim 2.5$ G and in the Hall limit for stronger fields.  

It is also clear that the degree of field -- matter coupling is strongly dependent on the abundance, and size distribution, of dust particles mixed with the gas. They reduce the fractional ionization as charged particles recombine on their surfaces and can also become an important species in high density regions (e.g.~Umebayashi \& Nakano 1990 \cite{1990MNRAS.243..103U}, Nishi et al. \cite{1991ApJ...368..181N}). As dust grains generally have large cross sections, they typically become decoupled from the magnetic field at lower densities than other --smaller-- species do. As a result of these processes, the conductivity of the fluid diminishes when dust particles (particularly when they are small) are well mixed with the gas (for example in early stages of accretion, or when turbulent motions prevent them from settling towards the disc midplane). Recent diffusivity calculations by Wardle 2007 \cite{2007Ap&SS.311...35W} for a minimum-mass solar nebula disc at 1 AU, have shown that  the magnetic coupling can in fact be adequate over the entire disc cross-section, once dust grains have settled. 

\section{Angular momentum transport}
\label{sec:L}

The mechanisms responsible for angular momentum transport in protostellar discs remain poorly understood. It is, however, increasingly clear that the two most promising processes are the following.

\begin{enumerate}

\item \emph{Vertical transport via winds and outflows} accelerated centrifugally from the disc surfaces (`disc winds'; e.g. see the reviews by K\"onigl \& Pudritz 2000 \cite{2000prpl.conf..759K} and Pudritz et al. 2007 \cite{2007prpl.conf..277P}), a notion strengthened by their ubiquity in star forming regions.

\item \emph{Radial transport through turbulent viscosity} induced by the magnetorotational instability (MRI; Balbus \& Hawley 1991 \cite{1991ApJ...376..214B} and 1998 \cite{1998RvMP...70....1B}). The MRI, essentially, taps into the free energy contributed by the differential rotation of the disc and converts it into turbulent motions that transfer angular momentum radially outwards.

\end{enumerate}

In the next two sections, we describe in more detail the operation of these two forms of transport in protostellar discs. We then briefly consider the possibility that they operate at the same radial location, but at different heights above the disc midplane.

 \subsection{Vertical transport via magnetocentrifugal outflows}
 \label{subsec:disk-wind}
 
 \begin{figure*}[]
\centering
\includegraphics[width=6.6in]{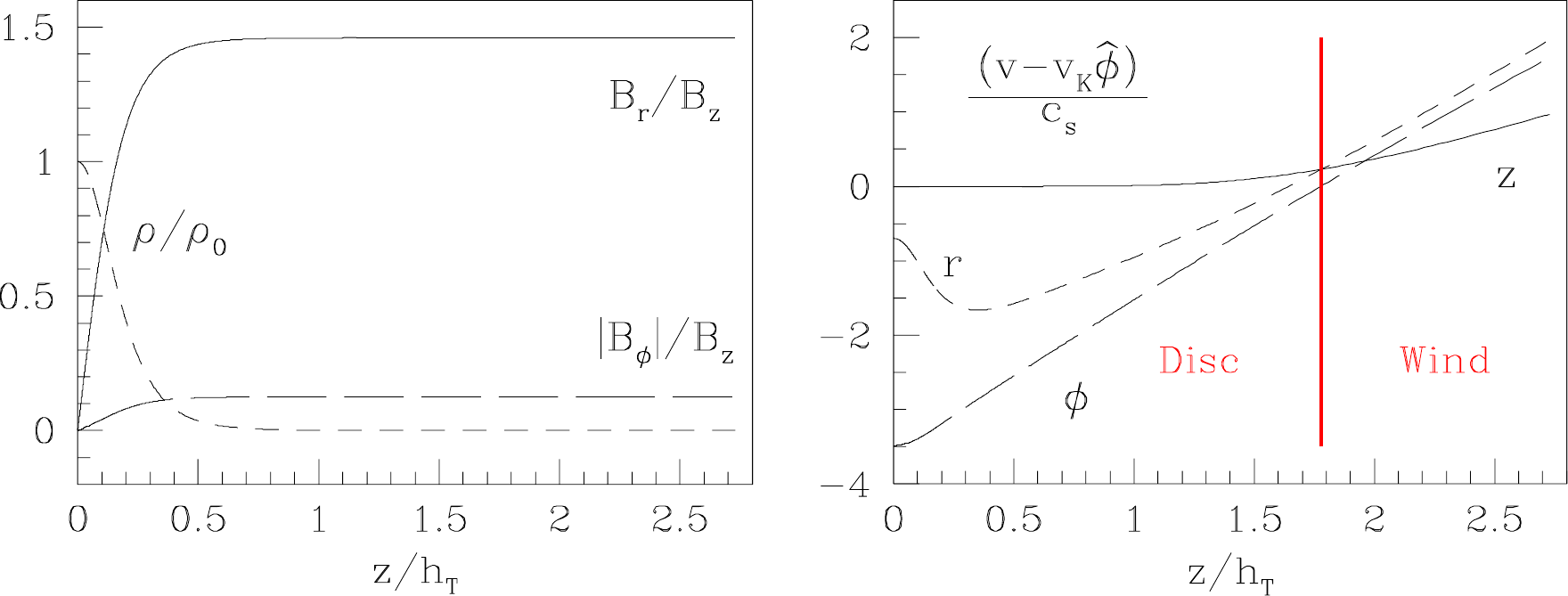}
\caption{Vertical structure of a strongly magnetised, wind--driving disc solution, as a function of height, under the assumption of pure ambipolar diffusivity. The model parameters are $a_0 = 0.95$,  $\eta = 10$, $\epsilon = 0.7$, $c_{\rm s}/v_{\rm K} = 20$ and $\epsilon_B = 0$ (see text). \emph{Left}: Normalised density ($\rho/\rho_{\rm 0}$) and magnetic field components (radial and azimuthal, $B_r/B_z$ and $|B_\phi|/B_z$, respectively, normalised by the vertical field $B_z$, which is constant with height). \emph{Right}: Velocity components, with respect to the Keplerian velocity $v_{\rm K}$ and normalized by the isothermal sound speed $c_{\rm s}$. Within the disc (to the left of the red vertical line) the radial velocity is negative, the azimuthal velocity is sub-Keplerian and the vertical velocity is small. In the outflow region, all three velocity components are positive.}
\label{fig:illus}
\end{figure*}

  \begin{figure*}[tbp]
\centering
\includegraphics[width=6.6in]{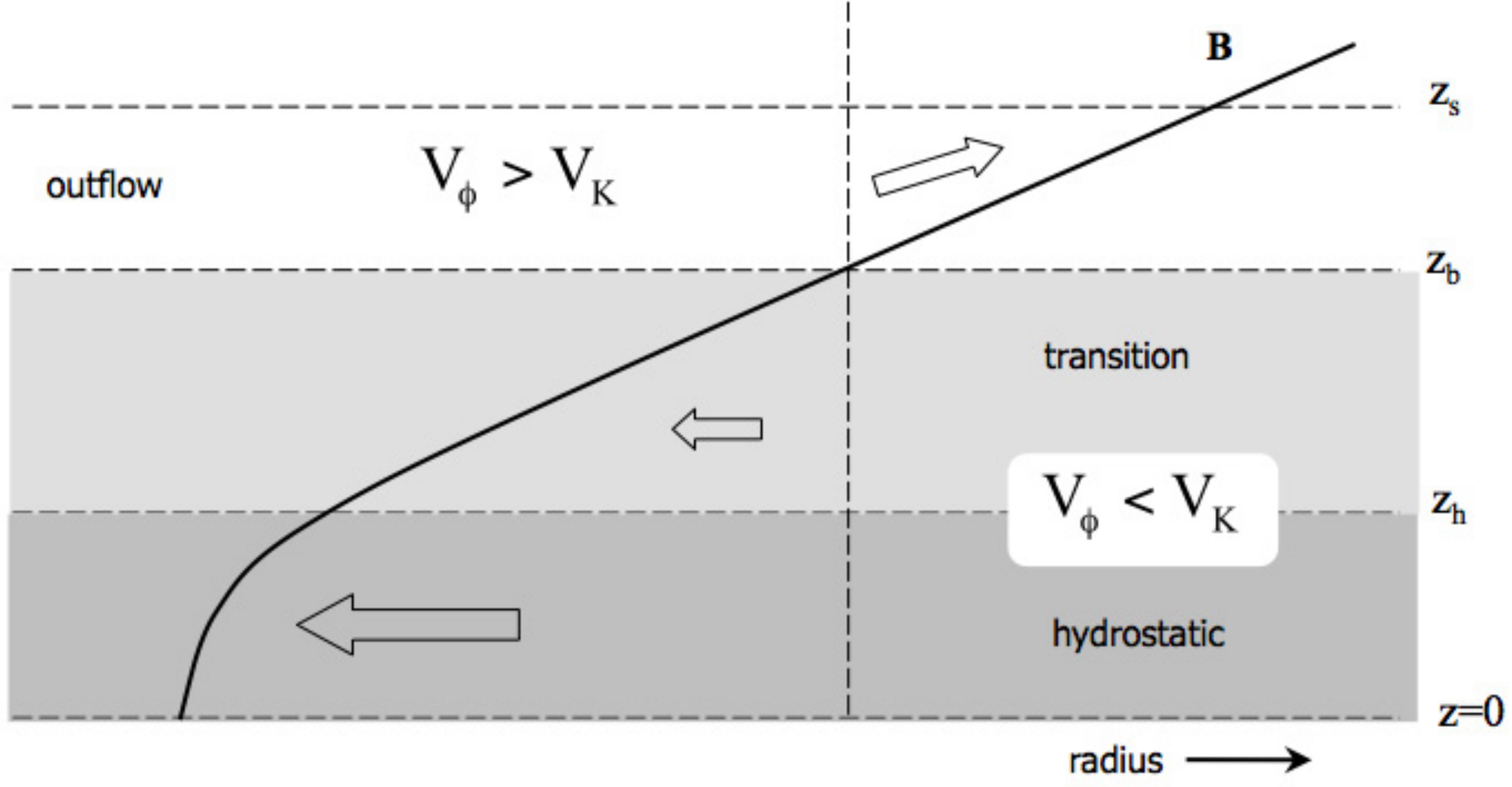}
\caption{Vertical structure of a wind-driving disc (Wardle \& K\"onigl 1993 \cite{1993ApJ...410..218W}). Three distinct layers can be identified. (1) The \emph{quasi-hydrostatic region} next to the midplane where most of the accreting matter is located; (2) a \emph{transition layer}, where the inflow diminishes; and (3) an \emph{outflow region}, which constitutes the base of the wind. Note that in the first two layers (i.e.~within the disc) the radial velocity is negative, the azimuthal velocity is sub-Keplerian and the vertical velocity is small. In the outflow layer all these velocities are positive. Figure adapted from Wardle \& K\"onigl (1993) \cite{1993ApJ...410..218W}.}
\label{fig:sim}
\end{figure*}


Fast, collimated winds are commonly observed in accreting astrophysical systems. Furthermore, accretion and outflow signatures in these objects seem to be correlated\footnote{Typically, accreting objects show excesses in the UV, infrared and millimeter emission; whereas outflow activity is associated with P-Cygni profiles, forbidden line emission and the presence of molecular lobes.} (Hartigan et al.~1995 \cite{1995ApJ...452..736H}.  These features suggest that there may be a physical mechanism linking the outflow and accretion processes in these systems (K\"onigl 1989 \cite{1989ApJ...342..208K}).  Furthermore, it is thought that these winds are magnetocentrifugally accelerated from the disc via the mechanism first expounded by Blandford \& Payne 1982 \cite{1982MNRAS.199..883B}. In the scenario proposed by these authors, matter is accelerated along the magnetic field lines if they are sufficiently inclined (i.e.~by more than $30^\circ$) with respect to the angular velocity vector of the disc. This ejection initiates an outflow that can, potentially, reach super Alfv\'enic speeds.

Fig.~\ref{fig:illus} shows a typical disc-wind solution, plotted as a function of $h_{\rm T}$ (the scale height associated with tidal gravitational compression). This solution has been computed using the procedure detailed in Salmeron, K\"onigl \& Wardle 2007 \cite{2007MNRAS.375..177S}, in turn based on the formulation by \cite{1993ApJ...410..218W}. The model parameters are:

\begin{enumerate}
\item  $a_0 = v_{{\rm A}0}/c_{\rm s}$, the midplane (subscript 0) ratio of the Alfv\'en
speed to the sound speed. This parameter measures the magnetic field strength; 
\item $\eta$, the ratio of the Keplerian rotation time to the neutral--ion momentum-exchange time, which determines the degree of coupling between the neutrals and the magnetic field (with $\eta \gg 1$ and $\ll 1$ corresponding to strong and weak coupling, respectively); 
\item $\epsilon \equiv -v_{r0}/c_{\rm s}$, the normalized inward radial
speed at the midplane (subscript 0). This parameter is evaluated by imposing the Alfv\'en critical-point constraint on the wind solution;
\item $c_{\rm s}/v_{\rm K} = h_{\rm T}/r$, the ratio of the disc tidal scale
height to the radius (e.g.~the disc geometric thickness); and  
\item $\epsilon_B \equiv -cE_{\phi}/c_{\rm s} B_z$, the normalized azimuthal
component of the electric field, which is nonzero if the magnetic field lines are allowed to drift
radially (WK93).\footnote{\label{ansatz}Setting 
$\epsilon_{\rm B} = 0$ effectively fixes the value of $B_r$ at the disc surface (subscript `b'). In a 
global formulation $B_r$ would be determined by the radial distribution of $B_z$ 
(e.g. Ogilvie \& Livio 2001 \cite{2001ApJ...553..158O}; Krasnopolsky \& K\"onigl 2002 \cite{2002ApJ...580..987K}). }
\end{enumerate}

The left panel of the figure displays the density (normalised by its value at the midplane) and the radial and azimuthal components of the magnetic field, normalized by the vertical field (which is assumed not to vary with height). The right panel shows all velocity components, measured with respect to the Keplerian velocity $v_{\rm K}$ and normalised by the isothermal sound speed $c_{\rm s}$. This solution illustrates the main features of winds accelerated centrifugally from strongly magnetised discs (Wardle \& K\"onigl 1993 \cite{1993ApJ...410..218W}). Three distinct layers can be identified, as detailed below (see \cite{1993ApJ...410..218W} and Fig.~\ref{fig:sim})
\begin{itemize}
\item The \emph{quasi--hydrostatic} layer, next to the midplane, which is matter dominated. In this layer, the topology of the magnetic field is consistent with the field lines being radially bent and azimuthally sheared. The field removes angular momentum from the gas, as a result of the ion-neutral drag, and matter moves radially inwards. Note that a higher radial inflow corresponds to a stronger drag, since the field remains pinned at the midplane under the assumption that $\epsilon_{\rm B} = 0$, and results in a stronger bending of the lines. In a generalization to a global treatment, the surface value of $B_r$ would be determined by the radial distribution of $B_z$ along the disc (see also footnote \ref{ansatz}).
\item The \emph{transition} region, above the hydrostatic layer, where magnetic energy dominates and the magnetic field lines are locally straight (note that they are still inclined outwards). In this zone (as in the quasi-hydrostatic layer below), the fluid azimuthal velocity is sub-Keplerian, as magnetic tension helps support the fluid against the gravitational pull of the central protostar. 
\item As the azimuthal velocity of the field lines increases with height, eventually they overtake the fluid in which they are embedded. This is the upper \emph{wind region}, the base of the outflow, where the field transfers angular momentum back to the matter and accelerates it centrifugally. In this region all velocity components are positive.
\end{itemize}

Solutions such as the one shown in Fig.~\ref{fig:illus} are found for strong fields (such that the ratio $a$ of the Alfv\'en speed to the isothermal sound speed is $\lesssim 1$). In these conditions, the MRI is suppressed because the wavelength of the fastest growing mode exceeds the magnetically compressed scale height. The particular solution in the figure also assumes that the magnetic field is strongly coupled to matter ($\eta \gg 1$, note also that it is constant with height) and that ambipolar diffusion dominates over the entire section of the disc. Under these approximations it is self-consistent to find relatively high values of the parameter $\epsilon$ (of the order of 0.3 -- 1). In a more real configuration it would be necessary to incorporate the vertical stratification of the diffusivity, with different regimes dominating at different vertical locations (see section \ref{subsec:Magdiff}), and possibly a magnetically weakly coupled region in the disc interior. In the latter case, in particular, more moderate values of the inward flow speed at the midplane are expected (see Li 1996 \cite{1996ApJ...465..855L} and Wardle 1997 \cite{1997ASPC..121..561W}).

\subsection{Radial transport via MRI-induced turbulence}
\label{subsec:MRI}

  \begin{figure}[tbp]
\centering
\includegraphics[width=3.2in]{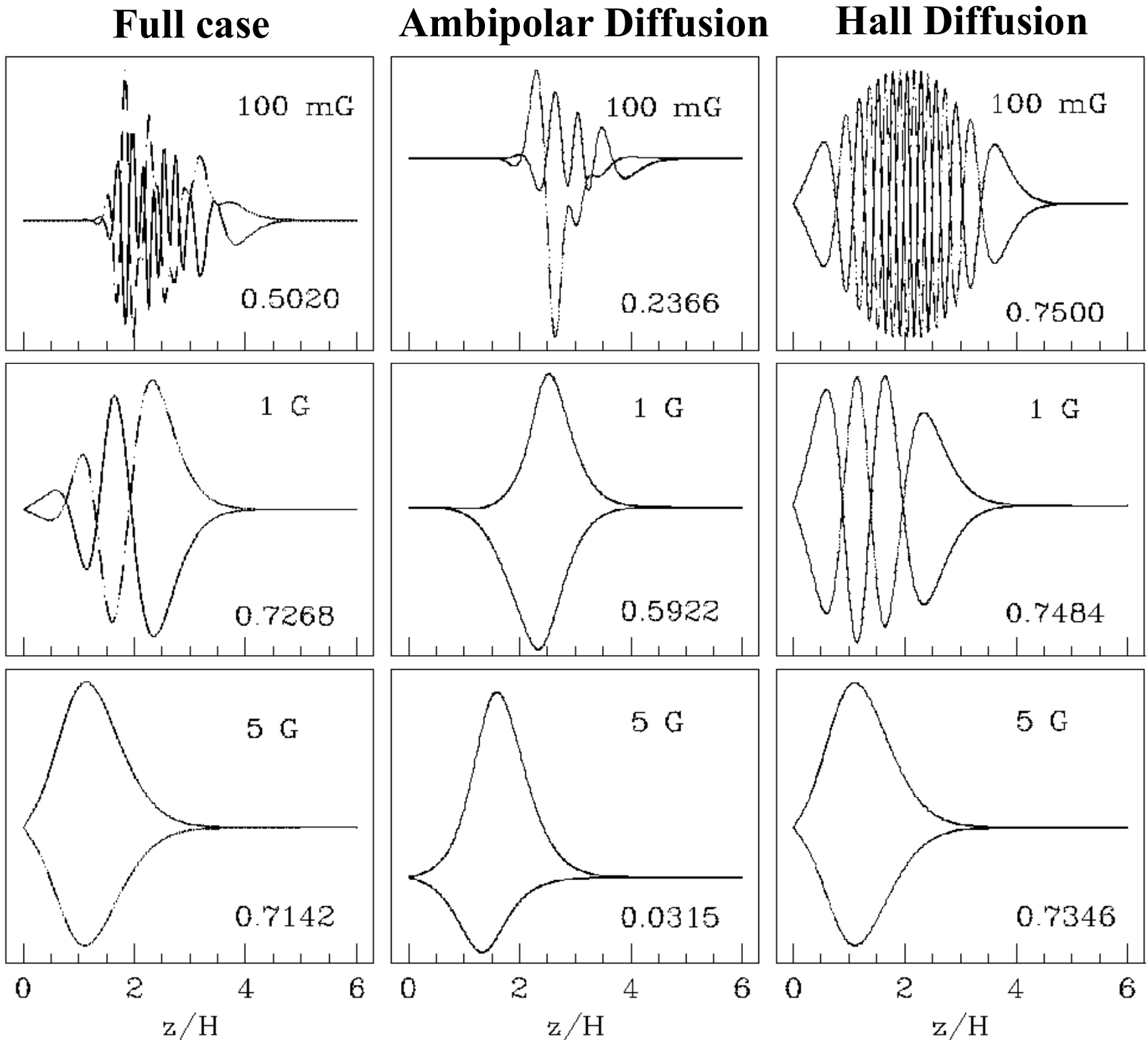}
\caption{Vertical structure and linear growth rate of the most unstable MRI modes for a minimum--mass solar nebula disc at $R = 1$ AU as a function of the magnetic field strength and for different diffusivity limits. In each panel, the solid (dashed) lines denote the radial (azimuthal) perturbations of the magnetic field, which is initially vertical. The middle and right columns show solutions obtained under the ambipolar and Hall diffusion approximations, respectively. The left column depicts solutions incorporating both diffusivity terms. The growth rate (in units of the Keplerian frequency) is indicated in the lower right corner of each panel. These results show that MRI perturbations grow faster, and are active closer to the midplane, when both diffusion mechanisms are considered. The velocity perturbations (not shown) are similarly affected (see also Fig.~2 and discussion in Salmeron, K\"onigl \& Wardle 2007 \cite{2007MNRAS.375..177S}). }
\label{fig:MRI1}
\end{figure}

  \begin{figure}[tbp]
\centering
\includegraphics[width=3.2in]{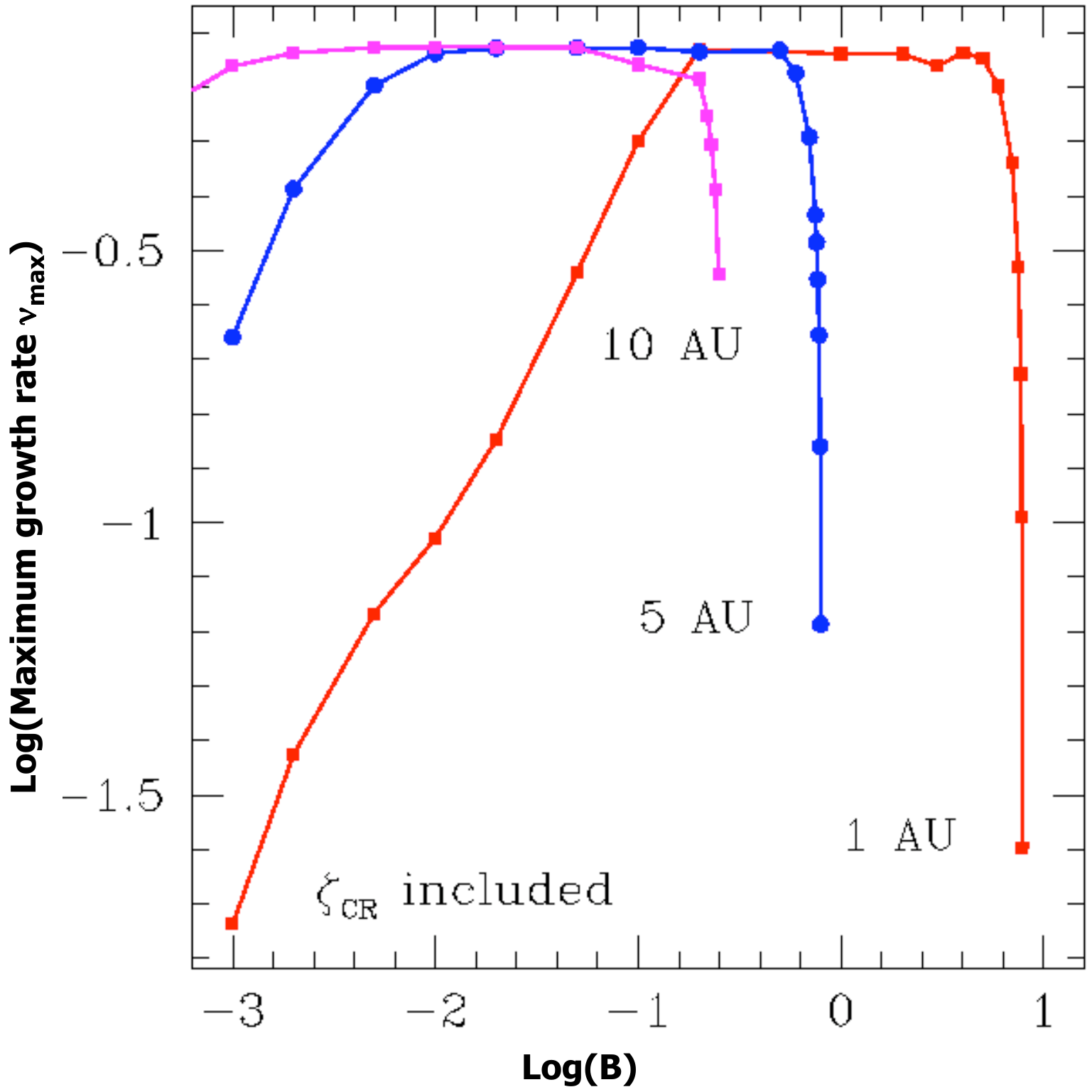}
\caption{Linear growth of the most unstable MRI modes for a minimum--mass solar nebula model at $R = 1$, $5$ and $10$ AU as a function of the strength of the magnetic field. }
\label{fig:MRI2}
\end{figure}

  \begin{figure}[tbp]
\centering
\includegraphics[width=3.2in]{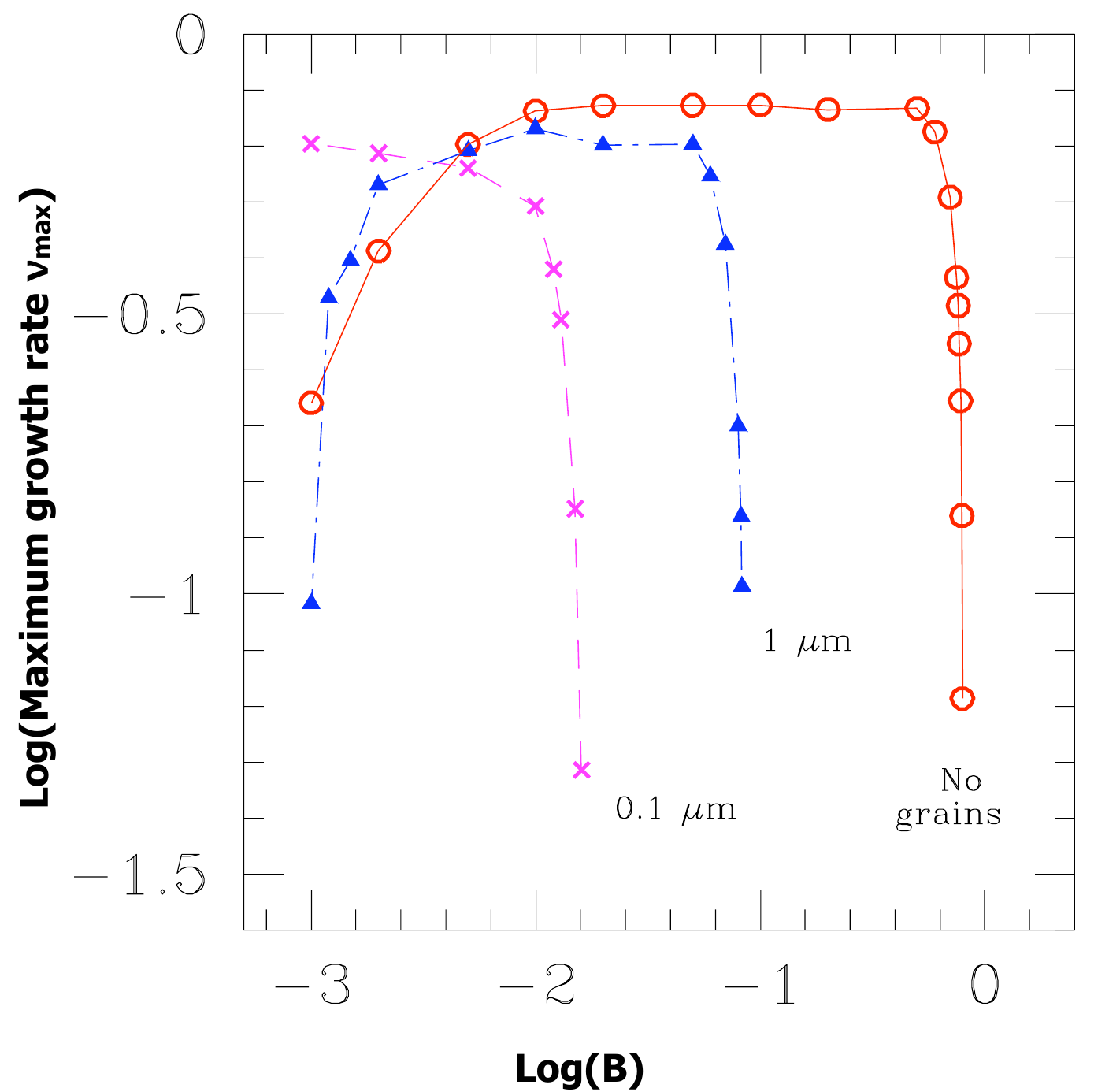}
\caption{As per Fig. \ref{fig:MRI2} for $R = 5$ AU and different assumptions regarding the presence, and size distribution, of dust particles suspended in the gas.}
\label{fig:MRI3}
\end{figure}

The MRI (Balbus \& Hawley 1991 \cite{1991ApJ...376..214B}, see also the review by these authors \cite{1998RvMP...70....1B}), transfers angular momentum radially outwards through the field lines that connect fluid elements at different radii. Under ideal-MHD conditions (negligible magnetic diffusivity), MRI-unstable modes grow in weakly magnetised discs (meaning here that the magnetic energy is small compared with the thermal energy density; or $a \ll 1$), provided that the angular velocity increases radially outwards. Because of the general conditions under which it operates, the MRI is a promising mechanism for the redistribution of angular momentum in astrophysical accreting systems.

Ideal-MHD conditions, however, are not likely to be satisfied in the deep, dense interiors of protostellar discs. In these environments, therefore, it is essential to consider how the low conductivity of the fluid affects the properties of the instability. Fig.~\ref{fig:MRI1} illustrates this point. It shows the vertical structure and linear growth rate of the fastest growing MRI perturbations ($\nu_{\rm max}$,  measured in units of the Keplerian frequency) for a minimum--mass solar nebula disc at $R = 1$ AU as a function of the magnetic field strength and for different diffusivity limits. In each panel, the solid (dashed) lines denote the radial (azimuthal) perturbations of the magnetic field, which is initially vertical. These solutions incorporate a realistic ionization profile and assume charges are carried by ions and electrons only (e.g.~dust grains have settled out of the gas phase).  They were obtained using the procedure detailed in Salmeron \& Wardle 2003 \cite{2003MNRAS.345..992S} and 2005 \cite{2005MNRAS.361...45S}, to which I refer the reader for additional details. Note that Hall diffusion strongly modifies the structure and growth of the modes, particularly when the magnetic field is weak. It is interesting that when Hall diffusion is included in the calculations, the vertical extension of the magnetically inactive zone next to the midplane is reduced. 

Fig.~\ref{fig:MRI2} compares the growth rate of the most unstable modes at different radii as a function of the magnetic field strength. Note that in all cases the growth rate initially increases with $B$, then levels at $\nu_{\rm max} \sim 0.75$ (the growth rate associated with ideal--MHD perturbations in a Keplerian disc) and, finally, drops abruptly at the field strength for which the wavelength of the perturbations is of the order of the tidal scaleheight. These results indicate that the MRI grows over a wide range of magnetic field strengths, even in the inner regions of weakly ionised discs. Note also that the maximum field strength for which perturbations grow is weaker at larger radii. This is expected, as the MRI is generally suppressed when the local ratio of the magnetic pressure to gas pressure (i.e. when 
the parameter $a$) 
is of order unity (Balbus \& Hawley 1991 \cite{1991ApJ...376..214B}). As both the gas density and the sound speed decrease with radius, the value of $a$ associated with a particular $B$ increases with $R$, causing the perturbations to be damped at a weaker field for larger radii.

Finally, the growth rate of the fastest growing MRI perturbations at $R = 5$ AU for different assumptions regarding the presence and size of dust grains mixed with the gas is shown in Fig.~\ref{fig:MRI3}.  Note that the range of field strengths over which the MRI operates is smaller as the grain size diminishes, as expected. However, the stabilizing effect of dust grains is not enough to completely damp the perturbations even when small particles ($0.1 \mu$m in size) are present. 

The results discussed above were obtained assuming that the surface density profile of the disc is that of the minimum-mass solar nebula model, $\Sigma(r) = 1700\  r_{\rm AU}^{-3/2}$ g cm$^{-2}$, where $r_{\rm AU}$ is the radial distance from the central object, measured in astronomical units. This yields a surface density of $\sim 150$ g cm$^{-2}$ at 5 AU and $\sim 50$ g cm$^{-2}$ at 10 AU. The surface density profile  in real discs may well be different from these inferred values. In fact, recent observations by Kitamura et al.~2002 \cite{2002ApJ...581..357K} and Andrews \& Williams 2007 \cite{2007ApJ...659..705A}, seem to imply a more gradual decline in surface density than what is expected from this model. In particular, the actual surface densities in the disc inner regions may be smaller than those of a minimum-mass solar nebula disc. This would facilitate a deeper penetration of the ionizing sources into the disc and would result in an increased ionization fraction closer to the midplane. This effect could modify the properties of MRI unstable modes in this region and, more generally, the presence and configuration of the magnetically dead zone assumed to exist in the disc interior (Gammie 1996 \cite{1996ApJ...457..355G}). On the other hand, protostellar discs may have surface densities roughly up to $\sim 50$ times the values implied by the minimum-mass solar nebula model without becoming gravitationally unstable. An increased surface density will result in a more extended section of the disc interior being shielded from the ionizing effect of X-rays and cosmic rays, but the magnetically active regions closer to the surface are not likely to be substantially modified.

 \subsection{Combined radial and vertical angular momentum transport}

The redistribution of angular momentum in real accretion discs is likely to involve \emph{both} the radial and vertical transport mechanisms described above. Salmeron, K\"onigl and Wardle (2007 \cite{2007MNRAS.375..177S}) have developed a novel methodology to
investigate the possibility that radial and vertical angular momentum transport occur at the same radial location (but at different heights) in real protostellar discs. Since the local value of $a$ increases with height on account of the strong decline in density (hence, gas pressure) away from the midplane, it is -- in principle -- possible for both forms of transport to operate at the same radius, with the MRI transporting angular momentum radially outwards over a section of the disc close to the midplane (where $a \ll 1$) and vertical transport operating at higher $z$ (satisfying $a \lesssim 1$). Preliminary results, computed in the limit where ambipolar diffusion dominates over the entire cross section of the disc, suggest that these two mechanisms may be radially segregated, an important finding with implications for the evolutionary paths of these discs by regulating the way matter is accreted (see details in Salmeron, K\"onigl and Wardle (2007 \cite{2007MNRAS.375..177S}).

\section{Conclusion}
\label{sec:planets}

We have examined the magnetic activity of protostellar accretion discs, focussing on the two leading mechanisms thought to be responsible for angular momentum transport in these objects: MRI--driven magnetic diffusivity and winds launched centrifugally from the disc surfaces.The first process redistributes angular momentum radially outwards via Maxwell stresses associated with small-scale (weak), disordered, magnetic fields. The second transfers it from the disc material to the wind through the action of a large scale (strong), ordered, field. In both cases, the low conductivity of the gas, particularly in the inner regions of the disc, has to be carefully accounted for when studying these processes. 
The results indicate that, despite the weak ionization, the field is able to couple to the gas and shear for fluid conditions thought to be satisfied over a wide range of radii in these discs.

\section*{Acknowledgment}

I thank the two referees for very useful comments that increased the clarity of the paper. 



\bibliographystyle{IEEEtran}
\bibliography{bibliography}		

\begin{thebibliography}{10}
\providecommand{\url}[1]{#1}
\csname url@samestyle\endcsname
\providecommand{\newblock}{\relax}
\providecommand{\bibinfo}[2]{#2}
\providecommand{\BIBentrySTDinterwordspacing}{\spaceskip=0pt\relax}
\providecommand{\BIBentryALTinterwordstretchfactor}{4}
\providecommand{\BIBentryALTinterwordspacing}{\spaceskip=\fontdimen2\font plus
\BIBentryALTinterwordstretchfactor\fontdimen3\font minus
  \fontdimen4\font\relax}
\providecommand{\BIBforeignlanguage}[2]{{%
\expandafter\ifx\csname l@#1\endcsname\relax
\typeout{** WARNING: IEEEtran.bst: No hyphenation pattern has been}%
\typeout{** loaded for the language `#1'. Using the pattern for}%
\typeout{** the default language instead.}%
\else
\language=\csname l@#1\endcsname
\fi
#2}}
\providecommand{\BIBdecl}{\relax}
\BIBdecl

\bibitem{2000prpl.conf..589S}
J.~M. {Stone}, C.~F. {Gammie}, S.~A. {Balbus}, and J.~F. {Hawley}, ``{Transport
  Processes in Protostellar Disks},'' \emph{Protostars and Planets IV}, pp.
  589--+, May 2000.

\bibitem{1973A&A....24..337S}
N.~I. {Shakura} and R.~A. {Syunyaev}, ``{Black holes in binary systems.
  Observational appearance.}'' \emph{\aap}, vol.~24, pp. 337--355, 1973.

\bibitem{1998RvMP...70....1B}
S.~A. {Balbus} and J.~F. {Hawley}, ``{Instability, turbulence, and enhanced
  transport in accretion disks},'' \emph{Reviews of Modern Physics}, vol.~70,
  pp. 1--53, Jan. 1998.

\bibitem{2000prpl.conf..759K}
A.~{Konigl} and R.~E. {Pudritz}, ``{Disk Winds and the Accretion-Outflow
  Connection},'' \emph{Protostars and Planets IV}, pp. 759--+, May 2000.

\bibitem{2007prpl.conf..277P}
R.~E. {Pudritz}, R.~{Ouyed}, C.~{Fendt}, and A.~{Brandenburg}, ``{Disk Winds,
  Jets, and Outflows: Theoretical and Computational Foundations},'' in
  \emph{Protostars and Planets V}, B.~{Reipurth}, D.~{Jewitt}, and K.~{Keil},
  Eds., 2007, pp. 277--294.

\bibitem{2006ApJ...647L..57S}
D.~{Semenov}, D.~{Wiebe}, and T.~{Henning}, ``{Gas-Phase CO in Protoplanetary
  Disks: A Challenge for Turbulent Mixing},'' \emph{\apjl}, vol. 647, pp.
  L57--L60, Aug. 2006.

\bibitem{2006AA...445..223I}
M.~{Ilgner} and R.~P. {Nelson}, ``{On the ionisation fraction in protoplanetary
  disks. II. The effect of turbulent mixing on gas-phase chemistry},''
  \emph{\aap}, vol. 445, pp. 223--232, Jan. 2006.

\bibitem{2006ApJ...639.1218T}
N.~J. {Turner}, K.~{Willacy}, G.~{Bryden}, and H.~W. {Yorke}, ``{Turbulent
  Mixing in the Outer Solar Nebula},'' \emph{\apj}, vol. 639, pp. 1218--1226,
  Mar. 2006.

\bibitem{2007ApJ...654L.159C}
F.~J. {Ciesla}, ``{Dust Coagulation and Settling in Layered Protoplanetary
  Disks},'' \emph{\apjl}, vol. 654, pp. L159--L162, Jan. 2007.

\bibitem{1996Icar..124...62P}
J.~B. {Pollack}, O.~{Hubickyj}, P.~{Bodenheimer}, J.~J. {Lissauer},
  M.~{Podolak}, and Y.~{Greenzweig}, ``{Formation of the Giant Planets by
  Concurrent Accretion of Solids and Gas},'' \emph{Icarus}, vol. 124, pp.
  62--85, Nov. 1996.

\bibitem{2003MNRAS.341.1157T}
C.~E.~J.~M.~L.~J. {Terquem}, ``{Stopping inward planetary migration by a
  toroidal magnetic field},'' \emph{\mnras}, vol. 341, pp. 1157--1173, Jun.
  2003.

\bibitem{2005MNRAS.363..943F}
S.~{Fromang}, C.~{Terquem}, and R.~P. {Nelson}, ``{Numerical simulations of
  type I planetary migration in non-turbulent magnetized discs},''
  \emph{\mnras}, vol. 363, pp. 943--953, Nov. 2005.

\bibitem{2008ApJ...679..813M}
T.~{Muto}, M.~N. {Machida}, and S.-i. {Inutsuka}, ``{The Effect of Poloidal
  Magnetic Field on Type I Planetary Migration: Significance of Magnetic
  Resonance},'' \emph{\apj}, vol. 679, pp. 813--826, May 2008.

\bibitem{2006ApJ...647.1413J}
E.~T. {Johnson}, J.~{Goodman}, and K.~{Menou}, ``{Diffusive Migration of
  Low-Mass Protoplanets in Turbulent Disks},'' \emph{\apj}, vol. 647, pp.
  1413--1425, Aug. 2006.

\bibitem{2005ASPC341..215C}
H.~C. {Connolly}, Jr., ``{Refractory Inclusions and Chondrules: Insights into a
  Protoplanetary Disk and Planet Formation},'' in \emph{Chondrites and the
  Protoplanetary Disk}, ser. Astronomical Society of the Pacific Conference
  Series, A.~N. {Krot}, E.~R.~D. {Scott}, and B.~{Reipurth}, Eds., vol. 341,
  Dec. 2005, pp. 215--+.

\bibitem{2003ApJ...585..908F}
T.~{Fleming} and J.~M. {Stone}, ``{Local Magnetohydrodynamic Models of Layered
  Accretion Disks},'' \emph{\apj}, vol. 585, pp. 908--920, Mar. 2003.

\bibitem{1977Ap&SS..51..153W}
S.~J. {Weidenschilling}, ``{The distribution of mass in the planetary system
  and solar nebula},'' \emph{\apss}, vol.~51, pp. 153--158, Sep. 1977.

\bibitem{1981Hayashi}
C.~{Hayashi}, ``{Structure of the solar nebula, growth and decay of magnetic
  fields and effects of magnetic and turbulent viscosities on the nebula},''
  \emph{Progress of Theoretical Physics Supplement}, vol.~70, pp. 35--52, 1981.

\bibitem{2004A&A...417...93S}
D.~{Semenov}, D.~{Wiebe}, and T.~{Henning}, ``{Reduction of chemical networks.
  II. Analysis of the fractional ionisation in protoplanetary discs},''
  \emph{\aap}, vol. 417, pp. 93--106, Apr. 2004.

\bibitem{2005ApJ...621..808G}
A.~E. {Glassgold}, P.~S. {Krsti{\'c}}, and D.~R. {Schultz}, ``{$H^{+}+H$
  Scattering and Ambipolar Diffusion Heating},'' \emph{\apj}, vol. 621, pp.
  808--816, Mar. 2005.

\bibitem{1999MNRAS.303..239W}
M.~{Wardle} and C.~{Ng}, ``{The conductivity of dense molecular gas},''
  \emph{\mnras}, vol. 303, pp. 239--246, Feb. 1999.

\bibitem{2002ApJ...570..314S}
T.~{Sano} and J.~M. {Stone}, ``{The Effect of the Hall Term on the Nonlinear
  Evolution of the Magnetorotational Instability. I. Local Axisymmetric
  Simulations},'' \emph{\apj}, vol. 570, pp. 314--328, May 2002.

\bibitem{Cowling1976}
{T. G. Cowling}, \emph{{Magnetohydrodynamics}}.\hskip 1em plus 0.5em minus
  0.4em\relax {London}: {Hilger}, {1976}.

\bibitem{1985A&A...147..247N}
C.~{Norman} and J.~{Heyvaerts}, ``{Anomalous magnetic field diffusion during
  star formation},'' \emph{\aap}, vol. 147, pp. 247--256, Jun. 1985.

\bibitem{1986MNRAS.218..663N}
T.~{Nakano} and T.~{Umebayashi}, ``{Dissipation of magnetic fields in very
  dense interstellar clouds. I - Formulation and conditions for efficient
  dissipation},'' \emph{\mnras}, vol. 218, pp. 663--684, Feb. 1986.

\bibitem{1999MNRAS.307..849W}
M.~{Wardle}, ``{The Balbus-Hawley instability in weakly ionized discs},''
  \emph{\mnras}, vol. 307, pp. 849--856, Aug. 1999.

\bibitem{2002ApJ...577..534S}
T.~{Sano} and J.~M. {Stone}, ``{The Effect of the Hall Term on the Nonlinear
  Evolution of the Magnetorotational Instability. II. Saturation Level and
  Critical Magnetic Reynolds Number},'' \emph{\apj}, vol. 577, pp. 534--553,
  Sep. 2002.

\bibitem{2003MNRAS.345..992S}
R.~{Salmeron} and M.~{Wardle}, ``{Magnetorotational instability in stratified,
  weakly ionized accretion discs},'' \emph{\mnras}, vol. 345, pp. 992--1008,
  Nov. 2003.

\bibitem{2005MNRAS.361...45S}
------, ``{Magnetorotational instability in protoplanetary discs},''
  \emph{\mnras}, vol. 361, pp. 45--69, Jul. 2005.

\bibitem{2008MNRAS.388.1223S}
------, ``{Magnetorotational instability in protoplanetary discs: the effect of
  dust grains},'' \emph{\mnras}, vol. 388, pp. 1223--1238, Aug. 2008.

\bibitem{1990MNRAS.243..103U}
T.~{Umebayashi} and T.~{Nakano}, ``{Magnetic flux loss from interstellar
  clouds},'' \emph{\mnras}, vol. 243, pp. 103--113, Mar. 1990.

\bibitem{1991ApJ...368..181N}
R.~{Nishi}, T.~{Nakano}, and T.~{Umebayashi}, ``{Magnetic flux loss from
  interstellar clouds with various grain-size distributions},'' \emph{\apj},
  vol. 368, pp. 181--194, Feb. 1991.

\bibitem{2007Ap&SS.311...35W}
M.~{Wardle}, ``{Magnetic fields in protoplanetary disks},'' \emph{\apss}, vol.
  311, pp. 35--45, Oct. 2007.

\bibitem{1991ApJ...376..214B}
S.~A. {Balbus} and J.~F. {Hawley}, ``{A powerful local shear instability in
  weakly magnetized disks. I - Linear analysis. II - Nonlinear evolution},''
  \emph{\apj}, vol. 376, pp. 214--233, Jul. 1991.

\bibitem{1993ApJ...410..218W}
M.~{Wardle} and A.~{Koenigl}, ``{The structure of protostellar accretion disks
  and the origin of bipolar flows},'' \emph{\apj}, vol. 410, pp. 218--238, Jun.
  1993.

\bibitem{1995ApJ...452..736H}
P.~{Hartigan}, S.~{Edwards}, and L.~{Ghandour}, ``{Disk Accretion and Mass Loss
  from Young Stars},'' \emph{\apj}, vol. 452, pp. 736--+, Oct. 1995.

\bibitem{1989ApJ...342..208K}
A.~{Konigl}, ``{Self-similar models of magnetized accretion disks},''
  \emph{\apj}, vol. 342, pp. 208--223, Jul. 1989.

\bibitem{1982MNRAS.199..883B}
R.~D. {Blandford} and D.~G. {Payne}, ``{Hydromagnetic flows from accretion
  discs and the production of radio jets},'' \emph{\mnras}, vol. 199, pp.
  883--903, Jun. 1982.

\bibitem{2007MNRAS.375..177S}
R.~{Salmeron}, A.~{K{\"o}nigl}, and M.~{Wardle}, ``{Angular momentum transport
  in protostellar discs},'' \emph{\mnras}, vol. 375, pp. 177--183, Feb. 2007.

\bibitem{2001ApJ...553..158O}
G.~I. {Ogilvie} and M.~{Livio}, ``{Launching of Jets and the Vertical Structure
  of Accretion Disks},'' \emph{\apj}, vol. 553, pp. 158--173, May 2001.

\bibitem{2002ApJ...580..987K}
R.~{Krasnopolsky} and A.~{K{\"o}nigl}, ``{Self-similar Collapse of Rotating
  Magnetic Molecular Cloud Cores},'' \emph{\apj}, vol. 580, pp. 987--1012, Dec.
  2002.

\bibitem{1996ApJ...465..855L}
Z.-Y. {Li}, ``{Magnetohydrodynamic Disk-Wind Connection: Magnetocentrifugal
  Winds from Ambipolar Diffusion-dominated Accretion Disks},'' \emph{\apj},
  vol. 465, pp. 855--+, Jul. 1996.

\bibitem{1997ASPC..121..561W}
M.~{Wardle}, ``{Magnetically-Driven Winds from Protostellar Disks},'' in
  \emph{IAU Colloq. 163: Accretion Phenomena and Related Outflows}, ser.
  Astronomical Society of the Pacific Conference Series, D.~T.
  {Wickramasinghe}, G.~V. {Bicknell}, and L.~{Ferrario}, Eds., vol. 121, 1997,
  pp. 561--+.

\bibitem{2002ApJ...581..357K}
Y.~{Kitamura}, M.~{Momose}, S.~{Yokogawa}, R.~{Kawabe}, M.~{Tamura}, and
  S.~{Ida}, ``{Investigation of the Physical Properties of Protoplanetary Disks
  around T Tauri Stars by a 1 Arcsecond Imaging Survey: Evolution and Diversity
  of the Disks in Their Accretion Stage},'' \emph{\apj}, vol. 581, pp.
  357--380, Dec. 2002.

\bibitem{2007ApJ...659..705A}
S.~M. {Andrews} and J.~P. {Williams}, ``{High-Resolution Submillimeter
  Constraints on Circumstellar Disk Structure},'' \emph{\apj}, vol. 659, pp.
  705--728, Apr. 2007.

\bibitem{1996ApJ...457..355G}
C.~F. {Gammie}, ``{Layered Accretion in T Tauri Disks},'' \emph{\apj}, vol.
  457, pp. 355--+, Jan. 1996.

\end{thebibliography}
%



\end{document}